# 30 GHz optoelectronic mixing in CVD graphene

A. Montanaro[*,1], S. Mzali[1], J.-P. Mazellier[1], O. Bezencenet[1], C. Larat[1], S. Molin[1], P. Legagneux[1], D. Dolfi[1], B. Dlubak[2], P. Seneor[2], M.-B. Martin[3], S. Hofmann[3], J. Robertson[3], A. Centano[4], A. Zurutuza[4]

[1]Thales Research and Technology, 1 Avenue Augustin Fresnel, 91767 Palaiseau
[2]Unité Mixte de Physique CNRS/Thales, 1 Avenue Augustin Fresnel, 91767 Palaiseau
[3]Department of Engineering, University of Cambridge, Cambridge CB21PZ, United Kingdom
[4]Graphenea S.A., TolosaHiribidea, 76 E-20018 Donostia, Spain

**Abstract** We report an optoelectronic mixer based on chemical vapour-deposited graphene. Our device consists in a coplanar waveguide that integrates a graphene channel, passivated with an atomic layer-deposited $Al_2O_3$ film. With this new structure, 30 GHz optoelectronic mixing in commercially-available graphene is demonstrated for the first time. In particular, using a 30 GHz intensity-modulated optical signal and a 29.9 GHz electrical signal, we show frequency downconversion to 100 MHz. These results open promising perspectives in the domain of optoelectronics for radar and radio-communication systems.

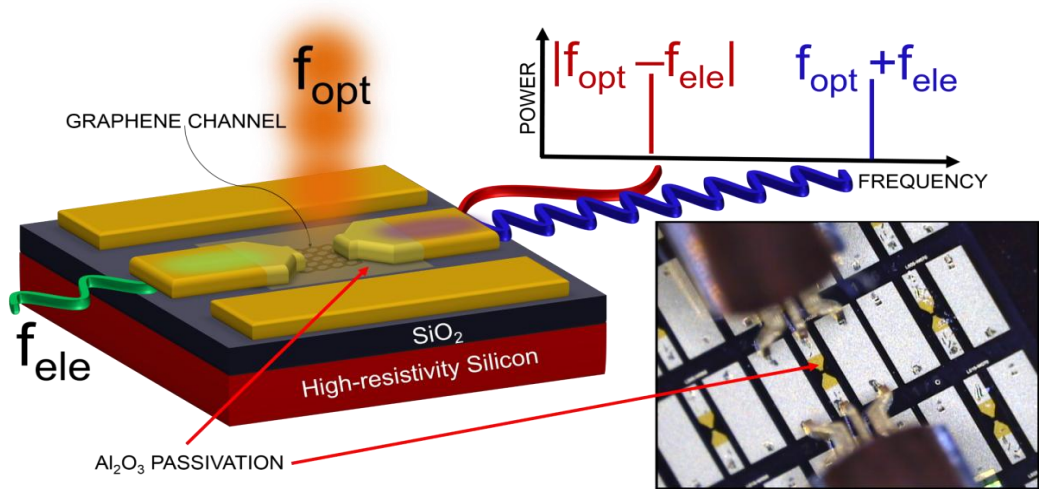

[*] Email: alberto.montanaro@thalesgroup.com

# Introduction

In high-speed communication systems, information transmission is carried on high-frequency (tens of GHz) signals. Once the information is received, a direct treatment at the carrier frequency is inconvenient, because it requires high-frequency electronic components. For this reason, a downconversion of the received signal is performed. Standard (low-frequency) electronics can then be used to treat the information. The downconversion function is commonly performed using mixers which demodulate high frequency carriers to base-band, by multiplying the carrier signal with a Local Oscillator (LO) signal. In many systems, one of the two signals is an optical one. In large Radio Detection And Ranging (RADAR) systems, for instance, the LO signal can be advantageously distributed to each antenna using an optical carrier. In this case, the LO signal has to be photodetected to be converted in the electrical domain. Then, the mixing is performed trough conventional electronic mixers. A more compact alternative is based on Optoelectronic Mixers (OEMs), which incorporate in the same device the photodetection and mixing function. More precisely, OEMs are devices that mix an electrical signal at frequency $f_{ele}$ and an intensity-modulated optical signal at frequency $f_{opt}$. Two electrical signals are generated at frequencies $f_{up} = f_{opt} + f_{ele}$ (up-conversion) and at $f_{down} = |f_{opt} - f_{ele}|$ (down-conversion). Performing simultaneously detection and mixing within the same device (self-mixing) reduces the number of components and potentially adds less noise in an optoelectronic system. All this features are particularly attractive for RADAR and Light Detection And Ranging (LIDAR) systems.[1-4].

State of the art OEMs that operate at 1.55 $\mu m$ wavelength are based on III-V semiconductors epitaxially grown on InP substrates. 60 GHz optoelectronic mixing has been demonstrated with Heterojunction phototransistors (HPT) based on InP-InGaAs,[5] and downconversion of 0.1THz signals has been performed by using Travelling-Wave Uni-Travelling carrier Photodiodes (TW-UTC-PD).[6] The research for low-cost and CMOS-compatible OEMs is active. 60 GHz bandwidth silicon-based OEMs operating at 0.85 $\mu m$ have been proposed.[7] However, optical communication systems are mainly based on 1.55 $\mu m$ band, and silicon is not suitable for efficient photodetection at this wavelength. Graphene can overcome the photodetection inefficiency of silicon, because it absorbs a broad spectrum of wavelengths,[8] including 1.55 $\mu m$.

Moreover, graphene exhibits very high carrier mobility and photocarrier short lifetime. Thanks to these remarkable properties, graphene is highly attractive in the domain of high frequency optoelectronics. Graphene-based photodetectors exhibit an intrinsic bandwidth that may exceed 500 GHz,[9] and CMOS-compatible graphene photodetectors have been demonstrated.[10] The combination of these characteristics, in junction with the strong efforts done for the realization of cost-effective production techniques for large-scale graphene,[11,12] makes this material a very good candidate for high-frequency low-cost OEMs.

Mao et al.[13] have fabricated the first graphene OEM based on a field effect transistor. In their work, the graphene channel is biased with a source-drain DC voltage and electrostatically doped by the gate voltage. They have demonstrated the mixing of a 2 MHz electrical signal with a 1 GHz intensity-modulated optical signal at 1.55 $\mu m$ wavelength.

Here we propose a new OEM structure that consists in a CVD graphene-based coplanar waveguide (gCPW). The graphene channel Fermi Level was set at the charge neutrality point (CNP), where the electron and hole concentrations are equal. No DC bias is applied to our device, to minimize power consumption. With this new structure, 30 GHz optoelectronic mixing in commercially-available graphene is demonstrated for the first time. In particular, we show a mixing between an intensity-modulated laser at 30 GHz and a 29.9 GHz electrical signal, and consequently we demonstrate frequency downconversion to 100 MHz.

## Fabrication and experimental setup

Figure 1 shows the operation principle of our graphene OEM and (on the bottom right side) a top view optical image of the device, contacted by RF electrical probes. We used a ground-signal-ground configuration, in which the central line (*signal line*) integrates a 23 $\mu m$ long and 20 $\mu m$ wide graphene channel.

To fabricate the device, a CVD graphene monolayer was first transferred on a high-resistivity silicon substrate covered by a 2 $\mu m$ thick thermal $SiO_2$ layer. Then, the graphene active zone was defined by optical lithography and reactive ion etching. A metallic multilayer (Ni-Ti-AuTi) was deposited by electron beam evaporation, and an optical lithography pattern allowed us to locally grow 2 $\mu m$ thick pads by electrolytic process, which reduces RF losses. The metallic multilayer deposited on the graphene channel was removed by successive etching techniques (dry/wet) until the Ni base layer was removed by wet chemistry. Finally, a 30 nm thick $Al_2O_3$ layer was deposited by Atomic Layer Deposition (ALD) to passivate the graphene channel.[14]

Figure 2 shows the experimental setup. We used a 1.55 $\mu m$ laser, modulated in intensity by a Mach-Zehnder Modulator (MZM). An Erbium-doped fiber amplifier (EDFA) amplified the modulated laser beam power, which was guided and focused on the CPW graphene channel. Two high frequency probes contacted the CPW ports. Both probes were connected to a bias tee to decouple the DC and the modulated components of the electrical signals, and allow DC biasing of each port. High frequency electrical signals where injected in the input port, while the output CPW port transferred the processed electrical signal to a spectrum analyzer. Finally, the high-resistivity silicon substrate was contacted and used as back-gate ($V_G$ in figure 2), to control the graphene Fermi-level with respect to the CNP.

## Results

Figure 3a shows the DC current that flows in the graphene channel of the device, as a function of the back-gate voltage, for a DC channel bias $V_{DC} = V_{IN} - V_{OUT} =$ 4V (see blue solid line in figure 3a). Thanks to the $Al_2O_3$ passivation layer, time-stable V-shaped curves are obtained[15]. The measurement allowed us to determine the CNP voltage ($V_{CNP}$). Then, the biased channel was illuminated by a laser modulated in intensity at a frequency $f_{opt}$ = 5 GHz. The modulated component of the optical power on the channel was $P_m$ = 22.5 mW. In the same figure (3a) the amplitude of the generated AC photocurrent ($I_{ph,m}$) is plotted as a function of the back-gate voltage (red dots). One can see that the photodetection has its maximal efficiency when the channel current is minimum. This behavior has already been observed by Freitag et al. in lightly doped graphene.[16] For this reason, all the following measurements have been performed at the CNP.

We performed two types of experiments. First, we characterized the device by maintaining a DC channel bias voltage (*photodetection*). Then, the constant bias was switched off and an RF signal was injected (*optoelectronic mixing*).

**Photodetection.** For the same laser parameters of figure 3a, figure 3b shows the amplitude of $I_{ph,m}$ as a function of the channel bias. The dependence is linear for voltages up to 6V. The $I_{ph,m}$ amplitude value obtained for $V_{DC}$ = 8V suggests that the photocurrent response starts to saturate for voltages above 6V. Such a behavior has already been observed.[16]

Figure 3c shows that the amplitude of $I_{ph,m}$ varies linearly with $P_m$ ($P_m$ varies between 0 and 22.5 mW). The modulation frequency was still $f_{opt}$ = 5 GHz and the DC channel bias was $V_{DC}$ = 6V.

For the same channel bias, the frequency response of the device is presented in figure 3d. Here, the photoresponsivity (namely, the photocurrent amplitude $I_{ph,m}$ normalized over the modulated component of the optical power $P_m$) is plotted as a function of $f_{opt}$ up to 30 GHz.

Figure 3b and 3c demonstrate that the photocurrent is a linear function of both the channel bias and the optical power. The photocurrent expression of a photodetector operating in such a linear regime can be written as:

$$I_{ph} = \alpha P_{opt} V_{bias} \qquad (1)$$

Where $P_{opt}$ is the optical incident power, $V_{bias}$ is the channel bias and $\alpha$ is a proportionality constant [$1/V^2$], which sizes the sensitivity of the device. In the experimental configuration described above, $V_{bias} = V_{DC}$ and $P_{opt} = P_{cw} + P_m sin(2\pi f_{opt} t)$, where $P_{cw}$ and $P_m$ are the amplitudes of, respectively, the constant and modulated components of the intensity-modulated laser beam, and $f_{opt}$ is the modulation frequency. We can thus rewrite equation 1 as:

$$I_{ph} = \alpha P_{cw} V_{bias} + \alpha P_m V_{bias} sin(2\pi f_{opt} t) = I_{ph,cw} + I_{ph,m} \qquad (2)$$

Where $I_{ph,cw}$ is the photocurrent component generated by the constant optical power $P_{cw}$, while the AC photocurrent (measured in plots 3b and 3c) generated by the modulated optical power is:

$$I_{ph,m} = \alpha P_m V_{DC} sin(2\pi f_{opt} t) \qquad (3)$$

Equation 3 expresses the AC photocurrent generated when the device is employed as a photodetector.

**Optoelectronic mixing.** In the optoelectronic mixing configuration, we switched off the constant voltage bias $V_{DC}$ and injected an RF modulated signal in the CPW, while maintaining the illumination of the graphene channel. The blue solid plot in figure 4a shows the power measured on a spectrum analyzer. The injected electrical power was 14 dBm, at frequency $f_{ele}$ = 400 KHz. $P_m$ was set to 22.5 mW, and $f_{opt}$ = 5 GHz. Two peaks appear at $f_{down}$ = |$f_{opt}$ − $f_{ele}$| = 4.9996 GHz and $f_{up}$ = $f_{opt}$ + $f_{ele}$ = 5.0004 GHz, experimentally demonstrating the optoelectronic mixing. One can notice that no signal at $f_{opt}$ is generated (no constant bias $V_{DC}$ is applied). For a direct comparison with the photodetection mode, in the same figure (4a), the red dashed curve shows the power signal when the modulated component of the electrical signal was switched off, and a constant $V_{bias}$ = $V_{DC}$ = 6V was applied.

For the optoelectronic mixing configuration, the channel bias in equation 2 is $V_{bias}$ = $V_m sin(2\pi f_{ele} t + \varphi)$, where $V_m$ is the amplitude of the RF electrical signal, and $\varphi$ is a phase shift with respect to the optical power modulation. $I_{ph,m}$ can thus be rewritten as:

$$I_{ph,m} = \alpha [P_m sin(2\pi f_{opt} t)][V_m sin(2\pi f_{ele} t + \varphi)] \tag{4}$$

giving:

$$I_{ph,m} = \frac{\alpha P_m V_m}{2}[cos(2\pi(f_{opt} - f_{ele})t - \phi) - cos(2\pi(f_{opt} + f_{ele})t + \phi)] \tag{5}$$

Due to the mixing between the electrical and optical signals, two signals are generated at frequencies $f_{up}$ = $f_{opt}$ + $f_{ele}$ and $f_{down}$ = |$f_{opt}$ − $f_{ele}$|. This explains the measurements. We then studied the optoelectronic mixing with high frequency (close to 30 GHz) electrical and optical signal carriers. In particular, for $f_{opt}$ = 30 GHz and $f_{ele}$ = 29.9 GHz, a downconversion to $f_{down}$ = 100 MHz is obtained (figure 4b).

Finally, we performed downconversion experiments for different electrical frequencies. An optical signal at frequency $f_{opt}$ = 10 GHz and $P_m$ = 22.5 mW was mixed with electrical signals at frequencies $f_{ele}$ varying from 6 to 9.99 GHz. Thus, the corresponding downconverted or intermediate frequency (IF) varied between 10 MHz and 4 GHz. We plotted the ratio between the downconverted electrical power ($P_{IF}$) and the electrical RF input power ($P_{IN}$). As can be seen, the downconversion response is "flat" as it oscillated (only) in a range of about 2 dBm over the entire frequency window.

# Conclusion

We studied a new high-frequency OEM which relies on a coplanar waveguide (gCPW) integrating a commercially-available CVD graphene channel. A simple model is used to describe its operating behaviour. This model is based on two linear dependencies, experimentally verified: the photocurrent as a function of the optical incident power ($P_{opt}$) at a fixed voltage drop along the channel, and the photocurrent as a function of the voltage drop ($V_{bias}$) along the channel at a fixed optical power. As the photocurrent is proportional to $P_{opt} \cdot V_{bias}$, upconverted and downconverted signals can be generated.

Used as a photodetector, our device exhibits an intrinsic photoresponsivity (i.e. after loss correction as explained in section *Methods*) that varies between 142 and 197 $\mu A$/W over a wide range of frequencies up to 30 GHz. Moreover, operating our gCPW as an optoelectronic mixer (OEM), we showed optoelectronic mixing at electrical/optical frequencies up to 30 GHz. We also showed a flat downconversion response (2 dB oscillation range) over a large frequency range. The demonstrated broadband downconversion seems very promising and opens interesting perspectives in the domain of low cost, CMOS compatible and high frequency OEMs.

# Methods

As presented in section *Fabrication and experimental setup*, the optical beam was modulated, amplified, and then focused on the device. Only the laser power impinging on the active area (graphene channel) was taken into account to calculate the incident power, while the part of the spot out of the graphene channel was excluded, because it didn't contribute to the generation of a photocurrent. We experimentally verified that no high frequency photocurrent was generated if there was no overlap between the laser spot and the graphene channel.

Plots 4a and 4b, show direct measurements on the spectrum analyser. All the other measurements took into account the device and cables losses. This correction has been done to extract the intrinsic response of graphene to light excitation. The correction method and the photocurrent and photoresponsivity calculation procedure is described in the following.

We first measured the device and cable electrical S parameters, from which we extracted the losses. For plots in figure 3a, 3b and 3c, the laser beam was modulated in intensity at a frequency $f_{opt}$ =5 GHz. We registered the electrical power value of the photodetected signal on the spectrum analyser and we subtracted the losses of the device at 5 GHz, as well as the cable losses between the output of the waveguide and the spectrum analyser. Finally, from this corrected power value, the photocurrent value was extracted taking into account the 50 Ω impedance of the spectrum analyzer. All the plotted photocurrent values are expressed in RMS. The same photocurrent extraction procedure has been adopted for plot 3d which shows the photoresponsivity ($\frac{I_{photo}}{P_m}$) at each registered frequency. The $P_m$ value was measured with an optical spectrum analyser. This value decreased with frequency due to the MZM bandwidth limitation (20 GHz). The same method was used to determine the power values plotted in figure 4c.

# Acknowledgment

This research was supported by the EU FP7 work programme under Grant GRAFOL (No. 285275) and Graphene Flagship (No. 604391).

# References

(1) G. Pillet, L. Morvan, D. Dolfi, and J.-P. Huignard. Wideband dual-frequency lidarradar for high-resolution ranging, profilometry, and doppler measurement. In SPIE Europe Security and Defence, pages 71140E–71140E. International Society for Optics and Photonics, 2008.

(2) W. C. Ruff, J. D. Bruno, S. W. Kennerly, K. Ritter, P. H. Shen, B. L. Stann, M. R. Stead, Z. G. Sztankay, and M. S. Tobin. Self-mixing detector candidates for an fm/cw ladar architecture, 2000.

(3) P. Ghelfi, F. Laghezza, F. Scotti, G. Serafino, A. Capria, S. Pinna, D. Onori, C. Porzi, M. Scaffardi, A. Malacarne, V. Vercesi, E. Lazzeri, F. Berizzi, and A. Bogoni. A fully photonics-based coherent radar system. Nature, 507(7492):341–345, Mar. 2014.

(4) V. Vercesi, D. Onori, F. Laghezza, F. Scotti, A. Bogoni, and M. Scaffardi. Frequencyagile dual-frequency lidar for integrated coherent radar-lidar architectures. Opt. Lett., 40(7):1358–1361, Apr 2015.

(5) C.-S. Choi, J.-H. Seo, W.-Y. Choi, H. Kamitsuna, M. Ida, and K. Kurishima. 60ghz bidirectional radio-on-fiber links based on inp-ingaas hpt optoelectronic mixers. Photonics Technology Letters, IEEE, 17(12):2721–2723, Dec 2005.

(6) E. Rouvalis, M. J. Fice, C. C. Renaud, and A. J. Seeds. Optoelectronic detection of millimetre-wave signals with travelling-wave uni-travelling carrier photodiodes. Opt. Express, 19(3):2079–2084, Jan 2011.

(7) H.-S. Kang and W.-Y. Choi. Cmos-compatible 60 ghz harmonic optoelectronic mixer. In Microwave Symposium, 2007. IEEE/MTT-S International, pages 233–236, June 2007.

(8) R. R. Nair, P. Blake, A. N. Grigorenko, K. S. Novoselov, T. J. Booth, T. Stauber, N. M. R. Peres, and A. K. Geim. Fine structure constant defines visual transparencyof graphene. Science, 320(5881):1308, 2008.

(9) F. Xia, T. Mueller, Y.-m. Lin, A. Valdes-Garcia, and P. Avouris. Ultrafast graphene photodetector. <u>Nat Nano</u>, 4(12):839–843, Dec. 2009.

(10) A. Pospischil, M. Humer, M. M. Furchi, D. Bachmann, R. Guider, T. Fromherz, and T. Mueller. Cmos-compatible graphene photodetector covering all optical communication bands. <u>Nat Photon</u>, 7(11):892–896, Nov. 2013.

(11) P. R. Kidambi, C. Ducati, B. Dlubak, D. Gardiner, R. S. Weatherup, M.-B. Martin, P. Seneor, H. Coles, and S. Hofmann. The parameter space of graphene chemical vapor deposition on polycrystalline cu. <u>The Journal of Physical Chemistry C</u>, 116(42):22492– 22501, 2012.

(12) A. C. Ferrari, F. Bonaccorso, V. Fal'ko, K. S. Novoselov, S. Roche, P. Boggild, S. Borini, F. H. L. Koppens, V. Palermo, N. Pugno, J. A. Garrido, R. Sordan, A. Bianco, L. Ballerini, M. Prato, E. Lidorikis, J. Kivioja, C. Marinelli, T. Ryhanen, A. Morpurgo, J. N. Coleman, V. Nicolosi, L. Colombo, A. Fert, M. Garcia-Hernandez, A. Bachtold, G. F. Schneider, F. Guinea, C. Dekker, M. Barbone, Z. Sun, C. Galiotis, A. N. Grigorenko, G. Konstantatos, A. Kis, M. Katsnelson, L. Vandersypen, A. Loiseau, V. Morandi, D. Neumaier, E. Treossi, V. Pellegrini, M. Polini, A. Tredicucci, G. M. Williams, B. Hee Hong, J.-H. Ahn, J. Min Kim, H. Zirath, B. J. van Wees, H. van der Zant, L. Occhipinti, A. Di Matteo, I. A. Kinloch, T. Seyller, E. Quesnel, X. Feng, K. Teo, N. Rupesinghe, P. Hakonen, S. R. T. Neil, Q. Tannock, T. Lofwander, and J. Kinaret.Science and technology roadmap for graphene, related two-dimensional crystals, and hybrid systems. <u>Nanoscale</u>, 7:4598–4810, 2015.

(13) X. Mao, C. Cheng, B. Huang, Z. Zhang, S. Gan, H. Chen, and H. Chen. Optoelectronic mixer based on graphene fet. <u>Electron Device Letters, IEEE</u>, 36(3):253–255, March 2015.

(14) M.-B. Martin, B. Dlubak, R. S. Weatherup, H. Yang, C. Deranlot, K. Bouzehouane, F. Petroff, A. Anane, S. Hofmann, J. Robertson, A. Fert, and P. Seneor. Sub-nanometer atomic layer deposition for spintronics in magnetic tunnel junctions based on graphene spin-filtering membranes. <u>ACS Nano</u>, 8(8):7890–7895, 2014. PMID: 24988469.

(15) A. A. Sagade, D. Neumaier, D. Schall, M. Otto, A. Pesquera, A. Centeno, A. Z. Elorza, and H. Kurz. Highly air stable passivation of graphene based field effect devices. <u>Nanoscale</u>, 7:3558–3564, 2015.

(16) M. Freitag, T. Low, F. Xia, and P. Avouris. Photoconductivity of biased graphene. <u>Nat Photon</u>, 7(1):53–59, Jan. 2013.

(17) T. Mueller, F. Xia, and P. Avouris. Graphene photodetectors for high-speed optical communications. <u>Nat Photon</u>, 4(5):297–301, May 2010.11

# Figures

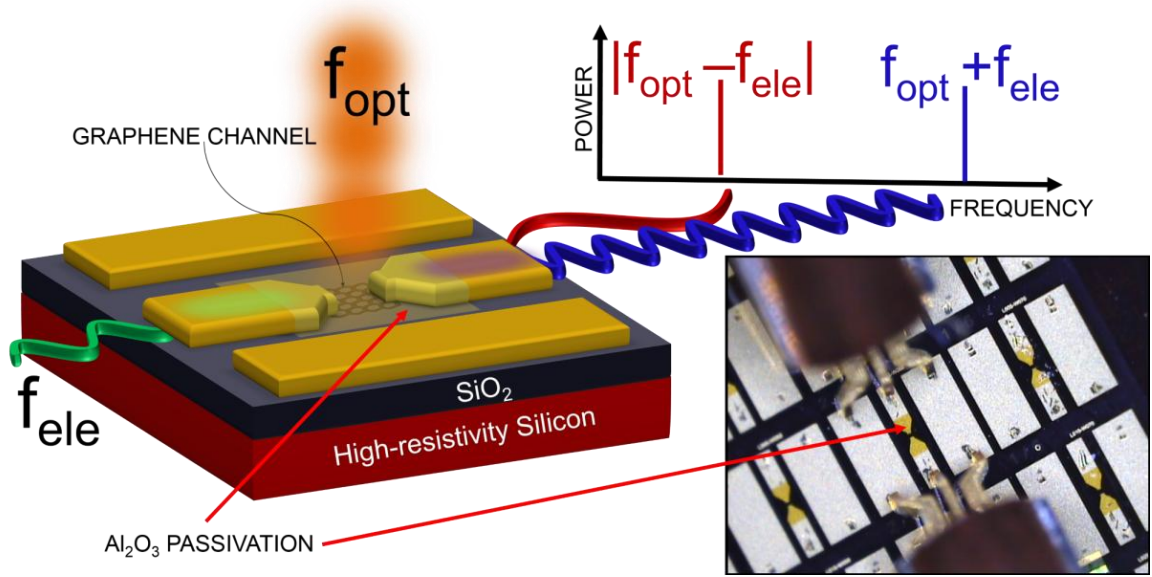

Figure 1: Schematic of our new graphene-based optoelectronic mixer (OEM). The simultaneous injection of an intensity-modulated laser beam at frequency $f_{opt}$ and an electrical RF signal at frequency $f_{ele}$ produces at the output two signals at the difference and sum of the input frequencies. *On the bottom right side*: Optical image of the CPW, contacted by the RF probes. The graphene channel is passivated with an $Al_2O_3$ layer. To allow the RF probes to electrically contact the device, the insulating film has been removed on the metallic pads.

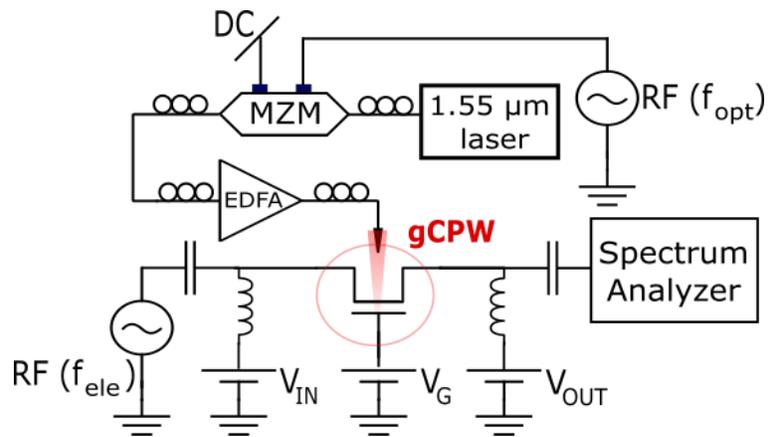

Figure 2: Experimental scheme. $V_{IN}$: DC Input Voltage; $V_{OUT}$ = DC output voltage; $V_G$ = Voltage applied on the high-resistivity silicon substrate, acting as back-gate. MZM: Mach-Zehnder Modulator; EDFA: Erbium-doped fiber amplifier; gCPW: graphene coplanar waveguide

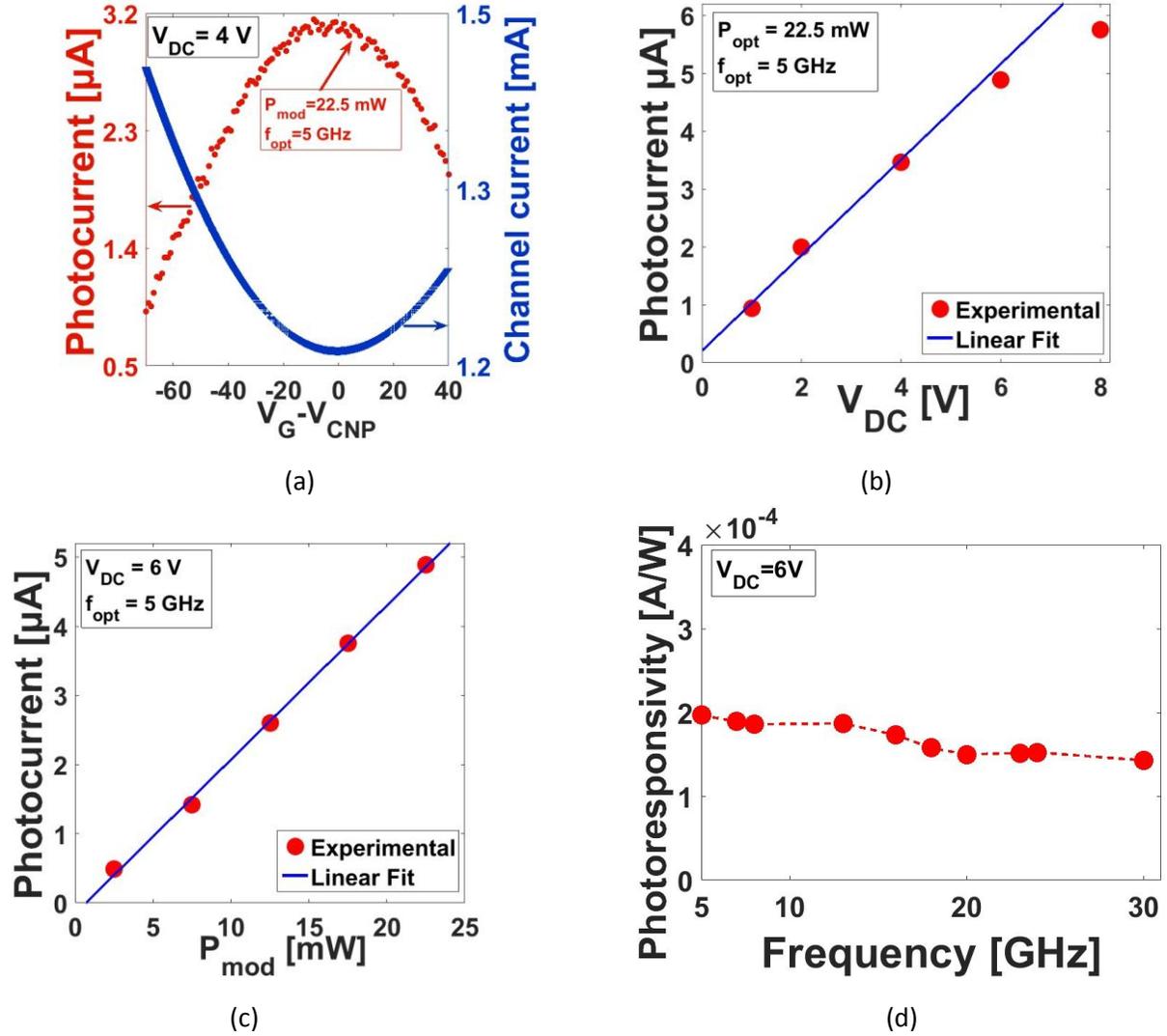

Figure 3: (a) *blue plot:* channel current as a function of $V_G - V_{CNP}$; *red plot*: photocurrent generated by a 5 GHz intensity-modulated laser beam. The modulated component of the laser beam is fixed at $P_m$= 22.5 mW. Both curves are measured under a channel bias voltage $V_{DC}$=4V. (b) Photocurrent as a function $V_{DC}$ for $P_m$ = $22.5 mW$ ($f_{opt}$ =5 GHz); the small residual current at 0 V bias predicted by the linear fit can be due to the nonperfect symmetry of the device.[17] (c)Photocurrent as a function of $P_m$ for $f_{opt}$ =5 GHz and $V_{DC}$ = $6V$. (d) Photoresponsivity ($I_{ph,m}/P_m$) as a function of the optical modulation frequency $f_{opt}$.
All the values are corrected with respect to the cables and device losses (see *Methods*).

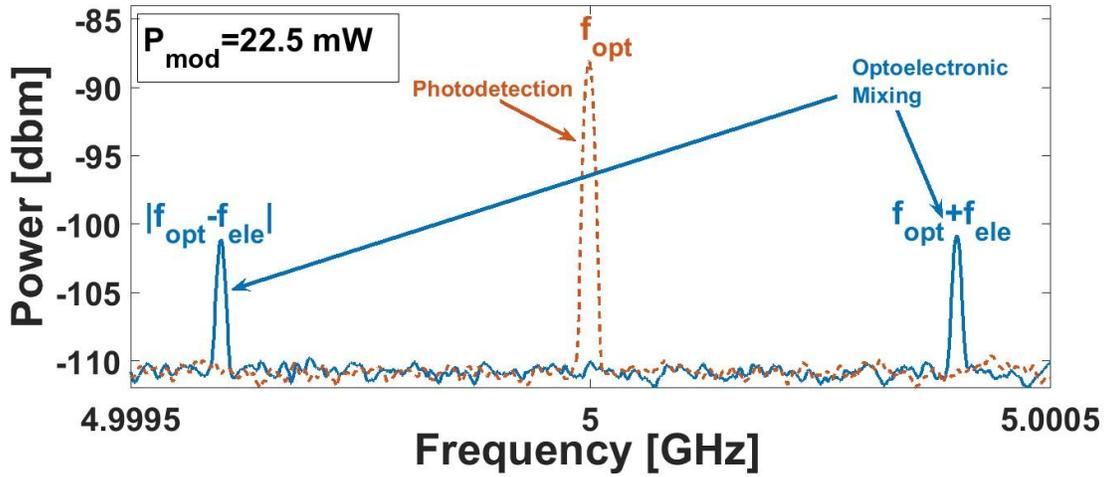

(a)

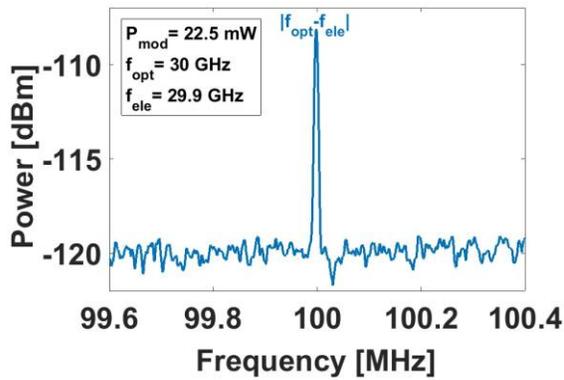

(b)

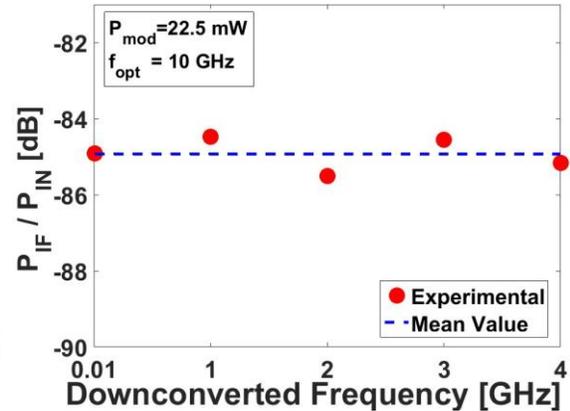

(c)

Figure 4: (a) *solid blue curve*: optoelectronic mixing effect, obtained by injecting a 14 dBm electrical signal at frequency $f_{ele}$=400 KHz. The laser parameters are: $P_m$ = 22.5 mW and $f_{opt}$ = 5 GHz. *Red dashed curve (for comparison)*: photodetection of the modulated component of the laser beam at 5 GHz. In this case, $V_{bias}$ = $V_{DC}$ = 6 V. (b) Downconversion power at 100 MHz of a laser signal at $f_{opt}$=30 GHz ($P_m$=22.5 mW), mixed with a 10 dBm electrical signal at 29.9 GHz.
(c) Optoelectronic mixing of a 10 GHz optical signal and an electrical signal at 6, 7, 8, 9, 9.99 GHz. The plot shows the power of the downconverted signal ($P_{IF}$) over the input electrical power ($P_{IN}$) as a function of the downconverted (or intermediate) frequency (IF). The laser power is $P_m$ = 22.5 mW.

15